\documentclass[12pt]{article}
\usepackage{times}
\usepackage{geometry}
\usepackage[utf8]{inputenc}
\usepackage{enumitem,amssymb}
\usepackage{ragged2e}
\newlist{thematic}{itemize}{8}
\setlist[thematic]{label=$\square$}
\usepackage{pifont}

\usepackage{newtxtext,newtxmath}
\usepackage{txfonts}
\usepackage{caption}
\usepackage[super,comma,sectionbib]{natbib}    
\usepackage{hyperref}                          
\hypersetup{colorlinks=true,linkcolor=blue}    

\usepackage{lscape}
\usepackage{longtable}
\usepackage{ulem}       
\usepackage[dvipsnames]{xcolor}
\usepackage[T1]{fontenc}
\usepackage{ae,aecompl}
\usepackage{graphicx}	
\usepackage{amsmath}	
\usepackage{amssymb}	
\newcommand{\aap}{A\&A}
\newcommand{\aj}{AJ}
\newcommand{\apj}{ApJ}
\newcommand{\apjl}{ApJ}
\newcommand{\apjs}{ApJS}
\newcommand{\araa}{ARA\&A}
\newcommand{\mnras}{MNRAS}
\newcommand{\nat}{Nature}
\newcommand{\pasp}{PASP}
\newcommand{\ssr}{Space Sci. Rev.}

\newcommand{\Msun}{$\rm \,  M_{\odot}$}
\newcommand{\Osun}{$\rm \,  O_{\odot}$}
\newcommand{\Fesun}{$\rm \,  Fe_{\odot}$}
\newcommand{\Zsun}{$\rm \,  Z_{\odot}$}

\newcommand{\Mini}{\mbox{$M_{ini}$}}
\newcommand{\Mstar}{\mbox{$M_{\ast}$}}

\newcommand{\Teff}{\mbox{$T_{\rm eff}$}}
\newcommand{\teff}{\mbox{$T_{\rm eff}$}}

\newcommand{\Lbol}{\mbox{$L_{\rm bol}$}}

\newcommand{\vsini}{$v {\rm sin}\, i$}

\newcommand{\lya}{$Ly \, {\alpha}$}
\newcommand{\Htwo}{$\rm H_2$}

\begin{document}
\raggedright
\LARGE

\Large


\hspace{-1cm}
\textbf{
Walking along Cosmic History: 
Metal-poor Massive Stars }  \\ 

\smallskip

  A White Paper Submitted to the Decadal Survey Committee A2020 \linebreak

\normalsize

\noindent \textbf{Thematic Areas:}

Stars and Stellar Evolution (primary) \\
Resolved Stellar Populations and their Environments (secondary)  \linebreak

\textbf{Principal Author:}

Name:	Miriam Garcia
\linebreak						
Institution: Centro de Astrobiolog\'{\i}a, CSIC-INTA
 \linebreak
Email: {\color{blue} mgg@cab.inta-csic.es}  \hspace*{60pt} 
Phone:  +34 91 520 2181
 \linebreak
 
 \textbf{Co-authors:}
 C.J. Evans (UK ATC, ROE), A. Wofford (IA-UNAM) \\
 J.C. Bouret (CNRS-LAM), N. Castro (AIP),
 M. Cervi\~no (CAB, INTA-CSIC), A.W. Fullerton (STScI),
 A. Herrero (IAC), D.J. Lennon (IAC), F. Najarro (CAB, INTA-CSIC)
  \linebreak

\vspace{-0.45cm}
 \justify      
  \textbf{Abstract:}
Multiple generations of massive stars have lived and died during Cosmic History,
  invigorating host galaxies with ionizing photons,
  kinetic energy, fresh  material
  and stellar-size black holes.
At present, massive stars in the Small Magellanic Cloud (SMC) serve as templates for
low-metallicity objects in the early Universe.
However, recent results have highlighted important differences in the
evolution, death and feedback of massive stars
with poorer metal content that better matches the extremely low metallicity of previous Cosmic epochs.
This paper proposes to supersede the SMC standard
with a new metallicity ladder built from very metal-poor galaxies,
and provides a brief overview of the technological facilities needed to this aim.

\begin{figure}[b!]
\centering
   \includegraphics[width=\textwidth]{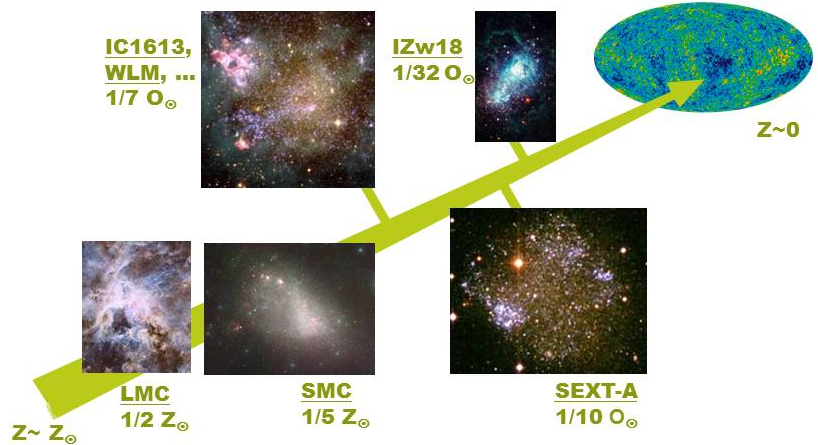}
   \label{F:road}
\vspace{-3cm}
\end{figure}

\pagebreak

\justify

\section{Introduction}  
\label{s:intro1}


Massive stars are stellar-size objects of the highest astrophysical impact for a number reasons.
Born with M$>$9\Msun~ they live fast and die spectacularly, making an excellent source of fast chemical
enrichment of the host galaxy.
At several stages of their evolution they experience very high effective temperatures
(\Teff$\geq$20,000~K, sometimes reaching up to 200,000~K) that results in an extreme ionizing
UV-radiation field.
The same UV-radiation can power supersonic winds that inject high amounts of kinetic
energy into the interstellar medium (ISM) and creates the gorgeous ionized bubbles
and complex HII structures often detected around massive stars.
The deaths of massive stars are counted among the most disrupting events ever registered:
type Ib,c,II supernovae (SNe), pair-instability SNe,
super-luminous supernovae (SLSNe) and long $\gamma$-ray bursts (GRB).
The surviving end-products, neutron-stars and stellar-size black holes, are sites of extreme physics. 
\textit{All these points render massive stars major agents of the Universe and galaxies,
whose inprint sprinkles many other fields of Astrophysics.}

Indeed, massive star feedback enters small and large-scale processes
spanning the age of the Universe, including the formation
of subsequent generations of stars (and planets) and the chemodynamical evolution of galaxies.
Since the Cosmic chemical complexity is ever-growing after the Big Bang,
the simulation and interpretation of these phenomena as we look back in time
demand robust theoretical predictions for massive stars 
at ever-decreasing metallicity.
Ultimately we need contrasted models for nearly metal-free very massive
stars that can be extrapolated to describe the First Stars of the Universe.

The massive stars of the Small Magellanic Cloud (SMC) constitute
the current standard of the metal-poor regime,
with an extensive battery of observations from ground- and space-based telescopes \citep{WFC02,Mal04,ELST06,LOSC16} 
providing evidence and constraints to inform theory.
All this is integrated into population synthesis codes
used to interpret observations of star-forming galaxies along Cosmic History.

However, the 1/5\Zsun~ metallicity of the SMC is not representative of
the Universe past redshift $z$=1 \citep{MD14}.
The theoretical framework for lower metallicities does exist \citep{MM02,Szal15,MGB17,EIT08}
and predicts substantial differences in the evolutionary pathways 
with impact in life-overall feedback and end-products.
We will elaborate further on this point, but we
highlight now that one of the proposed mechanisms to reproduce
the first gravitational wave ever detected
involves the binary evolution of two metal-poor massive stars \citep{AAA16,MM16}.

Teams around the world are working to build a representative
sample of massive stars with sub-SMC metallicity: a complete spectral atlas
that will allow us to draw the evolutionary pathways of massive stars,
and constrain their stellar properties (\teff, stellar mass \Mstar, luminosity \Lbol, and wind properties).
Unfortunately the SMC marks not only a metallicity but also a distance frontier,
and a sizeable leap down in metallicity requires reaching
distances of at least 1~Mpc (outer Local Group and surroundings).
Very promising galaxies with  1/10\Zsun~ (Sextans~A, 1.3~Mpc away) \citep{Cal16},
1/20\Zsun~ (SagDIG, 1.1~Mpc) \citep{G18} and 1/30\Zsun~ (Leo~P, 1.6~Mpc) \citep{ECG19}
are subject to close scrutiny with VLT, Keck and GTC
 \citep{Bal07,Eal07,Cal16,ECG19,GHN19},
with the \textit{holy-grail} 1/32 \Zsun~ starburst galaxy I~Zw18 (18.2~Mpc) \citep{VIP98,Aal07} always in mind.
However, the world's largest ground-based telescopes
only reach the brightest, un-reddened members after long integration times,
and even for these spectral quality is sometimes too poor as to yield
stellar parameters from quantitative analysis.
The result is a biased  and sorely incomplete view of sub-SMC massive stars.
\textit{The reality is that we have hit the limit of current observational facilities.}

This paper deals with metal-poor, sub-SMC metallicity massive stars.
The specifics of their evolution, interwoven with the evolution of galaxies and the Cosmos in the past,
has a high astrophysical relevance,
but the spectacular facts about their life and death are of interest to the general audience.
In the following pages we will summarize the state-of-the-art on the topic of metal-poor massive stars,
the exciting new scenarios that we may expect from the theoretical predictions,
how far we have reached with current observatories
and prospects for future missions in the planning.


\section{Stellar winds at low metallicity}

The evolution of massive stars is strongly conditioned by metallicity via stellar winds.
Massive stars experience very high effective temperatures \teff$>$20,000~K
during a great fraction of their evolution (Sect.~\ref{s:evol}).
The extreme UV radiation field 
exchanges energy and momentum with metal ions in the stellar atmosphere, resulting
in an outward outflow that we know as radiation-driven wind RDWs \citep{LS70,CAK75}.
The ensuing removal of mass \citep[$\sim 10^{-8} M_{\odot}/yr - 10^{-4} M_{\odot}/yr$][]{KP00}
is significant to
peel off the outer stellar layers (being responsible, for instance,
of the different flavours of Wolf-Rayet stars -WR-),
but also to alter the conditions at the stellar core and the rate of nuclear reactions.
It is because of the wind, which inherits a strong dependence on metal content
from its driving mechanism, that two massive stars born with the same initial
mass but different metallicity can follow very distinct evolutionary pathways
\citep{CM86} and result in different end-products (Sect.~\ref{s:evol}).

The winds of hot massive stas are weaker as their metallicity decreases.
The theoretical dependence of mass loss rate on metallicity
$ Mdot  \propto Z^{0.7-0.8}$~ \citep[][]{VKL01} has been verified
observationally down to the metallicity of the SMC \citep{Mal07}.
The winds of metal-poorer hot stars require a special formalism \citep{K02}
and the consideration that the driving ions shift from Fe to CNO at Z$\leq$0.1\Zsun,
and therefore the wind may evolve as processed material is brought to the surface by internal mixing.
The expectation is that at Z$< 10^{-2}$\Zsun~ winds are negligible, unless the star is very luminous,
and consequently will have very little impact on the evolution of the star.

Theory was finally confronted with observations with the arrival of multi-object spectrographs
at 8-10m telescopes.
The first efforts focused on IC~1613 (715~Kpc), the closest star-forming Local Group galaxy
whose $\sim$1/7\Osun~  nebular abundances marked a significant decrease 
in present-day metallicity from the SMC.
They soon were followed by studies in the $\sim$1~Mpc away galaxies NGC3109 and WLM.
The results were unanticipated:
the finding of an LBV with strong optical P~Cygni profiles \citep{HGU10a},
an extreme oxygen WR \citep{Tal13}, and the optical analysis of O-stars in the galaxies \citep{HGP12,Tal14}
all indicated that winds were stronger than predicted by theory at that metallicty.

Hubble Space Telescope (HST) played a crucial role deciphering the \textit{strong wind problem}.
The detailed analysis of UV spectral lines, more sensitive to the wind than the optical range,
yielded lower mass loss rates for O-stars \citep{Bal15}.
In parallel, 
UV spectroscopy showed that IC1613's content of iron was similar or even larger than the $\sim$1/5\Fesun~ content of the SMC,
superseding the 1/7\Fesun~ value scaled from oxygen
\citep{Gal14},
and similarly SMC-like Fe-abundances were also reported for WLM and NGC3109 \citep{Hal14,Bal15}.
While this finding
sets a reminder that metallicity cannot be scaled from oxygen abundances 
since the [$\alpha$/Fe] ratio reflects the chemical evolution of the host galaxy,
it also alleviates the discrepancy since the expected mass loss rate is larger
at the updated iron content \citep{VKL01}.

New efforts are being directed to the Sextans~A galaxy that has nebular abundances as low as
1/10-1/15\Osun~ \citep{Kal05} and similarly low stellar 1/10\Fesun~ abundances \citep{KVal04,Gal17}.
The first spectroscopic surveys 
have reported 16 OB stars \citep{Cal16},
but being located 1.32~Mpc away,
only 5 of them have been observed in the UV with HST-COS \citep{Gal17}.
The metal-poor component of the recently announced
\textit{ULLYSES} program (elaborated in response to HST's
\textit{UV Legacy DD Initiative})
will provide a priceless extension to such studies.
Only two other extremely metal-poor star-forming galaxies with resolved stellar population, have been
surveyed: SagDIG (1/20\Zsun, 1.1~Mpc), and Leo~P  (1/30\Zsun, 1.6~Mpc)
but they are either very far away or the foreground extinction severely hampers optical (let alone UV) observations.

Bottom-line is that the sample is insufficient and
the sub-SMC metallicity regime of RDWs remains unexplored.
It may be argued that the amount of mass lost to RDWs, specially at low metallicity,
may be negligible compared to other effects:
pulsation- and rotation-driven outflows, evolution and/or mass exchange in binary systems,
or eruptions such as those experienced by EtaCar \citep{S14}.
In fact, the concept that super-Eddington stars such as Eta Car may experience
continuum-driven winds, provides an interesting metallicity-independent mass loss mechanism \citep{vMOS}.  
These processes are very poorly understood compared to RDW even at solar metallicity, let alone among sub-SMC stars.
\textit{At the moment we simply lack any evidence to assess what is the dominant mass loss mechanism ruling the life of
metal-poor massive stars}.


\section{Evolution, explosions and feedback along Cosmic History}
\label{s:evol}

The close interaction between massive stars and the Universe began with the first generation of stars.
Primordial star formation simulations and
evidence from extremely metal-poor halo stars
are still working to settle 
the distribution of initial masses
\citep{HHY15,Br13,FCG17},
but at least a fraction of them were sufficiently massive and hot
as to commence the re-ionization of the Universe.
Ever since, signature of their copious ionizing flux
can be seen
in highly-ionized UV emission lines
(CIV1548,1551, OIII$]$1661,1666,
 $[$CIII$]$ 1907 + C III$]$ 1909) \citep{SSV17},
indirectly in Lyman-break galaxies -LBGs-,
and in a few interesting cases in the shape of \lya~emission -LAEs-,
that allows us to detect galaxies
and probe the cosmic star formation rate
out to redshift $z \sim$~10 (see e.g. introduction by \citep{WLS13}).

Therefore, understanding
massive stars with ever decreasing metal content matching
the composition of the Universe as we look back in time,
is the missing piece of information to interpret star-forming galaxies
in all their flavours: LAEs, LBGs, ULIRGs, Blue compact dwarfs, etc.
A proper characterization of the physical properties (\Teff, \Lbol, $Mdot$)
along their evolutionary stages, will enter population synthesis
and radiative transfer codes such as
Starburst99 \citep{LSG99} and CLOUDY \citep{FKV98},
to interpret observations of the integrated light from massive star populations \citep{WCB16}.
Armed with these tools to study the interplay between stars and hosts,
we can answer outstanding questions
such as 
the average ionizing photon escape fraction of galaxies, 
a crucial parameter to establish the end of the re-ionization epoch \citep{VOSH15,FPR12}.

The evolution of massive stars is more complicated to track
accross the Hertzsprung-Russell diagram (HRD) than their lower mass siblings.
They are born as O- or early B-dwarfs (\teff $\geq$ 20,000~K),
or extreme WNh when very massive.
After H-burning the star undergoes a sequence of evolutionary stages that strongly
varies with the initial stellar mass.
The zoo of post-MS stages includes 
O and B supergiants,
red supergiants -RSG-, luminous blue variables -LBV-, yellow hypergiants -YHG- and WR stars,
and covers an extreme temperature interval ranging from the $\sim$4000~K of RSG \citep{DKP13}
to $\sim$200,000~K in the most extreme oxygen WRs \citep{Tral15}.
Evolutionary models must link these stages, drawing paths that depend on 
metallicity, RDW mass loss and other outflows, rotational velocity or mass exchange in binary systems \citep{L12}.

Evolutionary tracks that properly deal with rotation and the
complicated physics of massive stars
have been extensively calculated for the Milky Way, LMC and SMC
\citep{Brott1,Ekal12},
Population III stars \citep{MCK03,EMC08,YDL12},
and for intermediate 1/50\Zsun~ metallicities \citep{Szal15}.
Significant changes are expected in the evolution of metal-poor massive stars,
some of them with tremendous impact on ionizing fluxes.
The most notable example is the incidence of chemically homogeneous evolution (CHE),
in which fresh He produced in the core
is brought to the surface by rotation-induced mixing.
A 1/5\Zsun, \Mini=25\Msun~ SMC star will usually reach the RSG stage,
but if the initial rotational velocity (\vsini) is extremely high it will evolve into
a CHE-induced
WR-like stage with \Teff$\sim$100,000~K \citep{Brott1}.
This effect is magnified at lower metallicities, where
very massive stars 
can either evolve into an envelope-inflated RSG,
or become a Transparent Wind Ultraviolet INtense star (TWUIN) \citep{Szal15}.
TWUINs double the HI ionizing luminosity and quadruple the HeII luminosity with respect to lower \vsini~ counterparts,
and could be responsible for the extreme HeII emission detected in I~Zw18 and the $z \sim $ 6.5 galaxy CR7, currently attributed to population~III stars \citep{Keal15,SMD15}.

Another fundamental aspect of the life of massive stars
is the end of their evolution.
There is a plethora of very energetic events associated to the death of massive stars:
core-collapse supernovae -SNe- (types Ib, Ic, II, IIL, IIn, IIP, IIb),
pair instability SNe, super-luminous SNe -SLSNe-, electron-capture SNe, hypernovae, kilonovae and long $\gamma$-ray bursts -LGRBs-.
Evolutionary models can predict the ending mechanism and leftover products
of single and binary systems \citep{WHW02,WH06,PWT17},
but observations have proven decisive to constrain and inform theory.
For instance, observations provided first evidence that RSG and LBV could explode as SNe
\citep{Gral13}; likewise,
the preference of LGRBs and SLSNe for metal-poor galaxies
\citep{LCB14,CSY17}
is a clue on the specific evolution of metal-poor massive stars.
Armed with a theoretical-sound, observationally-constrained \textit{map of progenitors} \citep{GY07,Sm09,vD17}
were the variation with metallicity is understood,
the most energentic LGRBs and SLSNe can be used to probe the high redshift Universe,
constrain star-formation rates \citep{Peal13} or even detect the signatures of the First Stars \citep{Br13}.

The LIGO and Virgo experiments have also revolutionized our view of massive
star evolution,
with 10 in-spiriling double black hole systems
detected during the short period of operations \citep{AA18}.
Numbers will soon enable statistics on the distribution of black holes -BH- and neutron star -NS-
masses, that will put the predicted scenarios for the fate of massive stars to the test.
Nonetheless LIGO and Virgo have already accomplished paradigm-shifting results.
The detection of GW170817, associated with a collapsing double neutron star 
and kilonova \citep{AAA17},
linked short-GRBs with massive stars  \citep{TPE17} in the greatest achievement of multi-messenger astronomy.
The very first gravitational wave system detected, GW150914, challenged
all we knew about the formation of black-hole systems.
With $36\:$M$_{\odot}$ and $29\:$M$_{\odot}$~ masses, the two BHs that merged
were significantly larger than all stellar-mass BHs that we know
($\sim$5-15\Msun) \citep{CJ14},
and those that could be formed from stellar evolution at solar metallicity
($\sim$ 20\Msun) \citep{SW14}.
This system has inspired the development of new scenarios, the most successful 
being the CHE evolution of two metal-poor massive stars that evolve
within their Roche Lobes avoiding mass exchange and the common-envelope phase \citep{MM16}.

Constraining massive stars evolution is a multi-dimensional problem,
even if focusing on metal-poor environments only.
The high incidence of massive star in multiple systems, and the fraction that
will interact with their companions \citep{Sal12},
complicates the problem exponentially.
The way to proceed is to assemble large, multi-epoch samples of massive stars to fully cover the parameter space,
reconstruct the evolutionary pathways,
constrain their physical properties
with the most advanced stellar atmosphere models,
obtain distributions of \vsini~ and of the properties of binary systems,
and contrast against the predictions of single and binary evolutionary models.
Such ensembles have been built over the years in the Milky Way \citep{SD15},
and most recently in the Magellanic Clouds \citep{Eal11,COFL18b}.
However, only a handful of massive stars have been confirmed by spectroscopy in galaxies with poorer metal-content than the Small Magellanic Cloud \citep{Gal17}.
At this stage no signature of CHE has been reported in these galaxies, and very few massive binaries are known.

\section{Massive star formation in metal-poor environments}
\label{s:imf}

How massive stars form remains a matter of intense investigation.
Our understanding of this topic has significant gaps
ranging from the formation of individuals,
to how the upper initial mass function (IMF) builds,
and whether there is a dependency on environment.
Two principal issues make the process markedly different from lower-mass sibblings:
the star-forming clumps must be prevented from breaking into smaller pieces,
and radiation pressure from the forming star may halt accretion.

Two main theories of star formation are emerging,
competitive accretion (radiation pressure is overcome by
  forming massive stars in the gravitational well of the whole cluster, including the possibility of mergers) \citep{BBC97}
and monolithic collapse (radiation is liberated via a jet, and results in one star per cloud) \citep{Kral09}.
At solar metallicity, they both struggle to form stars more massive than 20-40\Msun~ \citep{ZY07,HC14,TBC14}
in stark contrast with the number of known $\geq$60\Msun~ stars
in Milky Way clusters
(e.g. Westerlund~1 \citep{CNC05}, Carina \citep{S06},
Cygnus~OB2 \citep{BHC18} or the Galactic Center \citep{NdLFG17})
and growing evidence of a top-heavy initial mass function in resolved, bursty star forming regions
\citep{SSE18}.

The feasibility  of both scenarios eludes observational confrontation
since it is hard to catch forming massive stars in the act of formation.
The most massive young stellar objects (MYSOs) also have masses of 20-30\Msun~ \citep{NSF18},
but at this stage a highly embedded hot core is detected where it is complicated to disentangle
the contribution of the accretion structure and the ionized gas component
with both imaging and spectroscopy \citep{SLC09}.
Few interesting missing links exist like \textit{IRAS 13481-6124}, a 18\Msun~ MYSO
for which VLTI-AMBER detected
a 20\Msun~ surrounding disk \citep{KHM10}.
Nonetheless, this system only make $\sim$40\Msun~ at maximum efficiency.
There is still no evidence of a forming, single $\gtrsim$60\Msun~ star
and the number of candidate merger events \citep{SAV16}
or massive stars that result from mergers \citep{Vb12}
is too small at the moment.

The situation should be alleviated in environments
of decreasing metallicity, since the paucity of metals would both
prevent gas from cooling and breaking down into
smaller pieces,
and 
make pre-stellar radiation-driven outflows weaker \citep{V18}.
The former argument is fundamental to support the widely-accepted concept that
the \textit{first, metal-free, stars of the Universe were very massive}.
In fact, the record holder $\sim$150\Msun~ massive stars have been found
at the heart of the \textit{Tarantula Nebula} in 
the LMC \citep{Cal10} and have 0.4\Zsun~ metallicity.
Evidence of over 100\Msun~ stars has also been found in the integrated light of
unresolved, metal-poor starburst \citep{WLCB14,SCC16}

The sequence drawn by Local Group dIrr's
enables us
to investigate whether the upper mass limit is set to higher values as metallicity decreases,
reaching values similar to those at the peak of star formation of the Universe and beyond in the past \citep{MD14}:
IC1613, WLM and NGC3109 (1/7\Osun), Sextans~A (1/10\Osun), SagDIG (1/20\Osun) and Leo~P (1/30\Osun).
Some of these galaxies host spectacular HII shells equivalent in size to the 30~Dor region,
but no analog to the
LMC's \textit{monster stars} has been found yet.
The most massive star reported has an initial mass of only 60\Msun,
and only a handful of them have masses in the 40-60\Msun~ range \citep{Gal17}. 

It may be argued that other factors may outweigh the paucity of metals
such as the galactic reservoir of gas, the 
local gas density or the star-formation rate \citep{GHS11},
expecting higher mass stars in denser, more active regions.
This hypothesis was also challenged by the spectroscopic
detection of OB-stars at the low stellar and gas density outskirts of Sextans~A,
2 of them being the youngest and most massive stars known so far in the galaxy \citep{GHN19}.
Previous indications of young massive stars in low gas-density environments
existed, like the extreme UV-disk galaxies \citep{GdP05},
but O-stars identified spectroscopically enable the unequivocal association of
UV emission in the outskirts with young massive stars.


At this stage one may wonder how molecular gas behaves but
this important piece of the puzzle is missing.
Direct observations of cold \Htwo~ are unfeasible at 
most of the sub-SMC metallicity Local Group dIrr's (0.715 - 1.3~Mpc)
and CO is extremely challenging, because of the low metallicity.
CO has been detected only
in a few of them \citep{ERH13,SWZ15},
overlapping the highest concentrations of stars and UV emission,
but the low-density outskirts of the galaxies have not been targeted.
The mass of molecular gas can also be estimated from 
the dust content but at $\sim$1~Mpc Herschel may miss
the unconspicuous regions (see e.g. selection by \citep{SAH14})
and results rely heavily on the adopted gas to dust ratios
which in turn depend on metallicity with a large scatter \citep{GLK17}.
At the moment, there is no reliable inference of the distribution
of molecular gas in the dwarf irregular galaxies of the Local Group.

A tantalizing alternative is that 
star formation could proceed directly from neutral gas.
Simulations have shown that low-density, metal-poor neutral gas can reach sufficiently low temperatures
to proceed to star formation without forming \Htwo~ molecules \citep{Kr12},
breaking the association of star formation and molecular gas.
Interestingly, there is a strong spatial correlation between
the HI maps and the location of OB stars and associations
in the dwarf irregular galaxies of the Local Group \citep{GHC10}.
Is it possible that cloud fragmentation and star formation
follows different mechanisms in dense environments hosting molecular clouds,
and sparse, neutral-gas dominated regions?
This would be a natural explanation for the occurence of SLSNe in the outskirts of galaxies \citep{LCB14}.
If this concept was demonstrated, the simulations of the evolution of galaxies
would need to be revisited to check the significance of the missing
mass formed in low density dwarf galaxies and the outskirts of spirals.

Hence, the latest results not only highlight our poor understanding on massive star formation but also open new questions.
The joint study of resolved massive stars
and detailed maps of neutral and molecular gas,
will help to unravel the relative role played by \Htwo~ and HI in star formation,
whether it changes with galactic site, and whether it translates into different mechanisms 
that populate the IMF and the distribution of initial rotational velocity and binaries distinctly.
Ideally, untargeted, unbiased spectroscopic censuses of massive stars in clusters and galaxies would enable reconstructing
these distributions that are so important to understand the star formation, establishing the local upper mass limit in particular, and checking for any dependence on metallicity and gas content.
However, observations of the required spectral quality are out of reach to current technology because:
1) main sequence O-stars at $>$ 1~Mpc are at the sensitivity limit of optical spectrographs installed at 10m telescopes;
2) the program should include near-IR observations to overcome internal extinction (significant in dIrr galaxies) \citep{GHN19} and to reach MYSO, but both are bluntly out of reach to current IR spectrographs; and
3) the densest concentrations of gas an stars should be inspected to look for very massive stars and 30~Dor $\sim$150\Msun~ analogs, but these regions are hardly resolved by ground-based spectroscopic facilities even using adaptive optics.
Nonetheless such a phenomenal database would enable studying on-going star formation with unprecedented detail, and
possibly to re-calibrate photometric-based star formation indicators after properly accounting for the missing mass.





\section{Technological needs for a significant break-through}

Our partial understanding of metal-poor massive stars jeopardizes
the interpretation of SNe and LGRBs,
star-forming galaxies along Cosmic History,
and the re-ionization epoch.
Four questions summarize the next great challenges faced by the field:


$\bullet$~ Is the IMF universal? What is the upper mass limit?
Does it increase with decreasing metallicity?

$\bullet$~ What kind of outflows do metal-poor massive stars experience?

$\bullet$~ How do they evolve? What is the frequency of CHE?

$\bullet$~ What are the evolutionary channels that lead to binary stellar mass
  black holes?

Homogeneous analyses of high-quality optical and UV datasets
have provided invaluable insight into
massive stars of the Milky Way and the Magellanic Clouds \citep{SD15,Eal11,COFL18b}.
Notably, \textit{ULLYSES} is devoting a significant number of orbits
to ensure proper UV spectroscopic coverage of the SMC.
These and other on-going efforts will consolidate our knowledge
of massive stars at the present-day, and
lay the groundwork for the kind of in-depth studies
needed to provided quantitative results. 

Answering the stated questions requires
a phenomenal observational ensemble in 
sub-SMC metallicity galaxies
enabling an equivalent level of detailed work,
but this is unfeasible with current facilities.
The wavelength coverage, sensitivity and spatial resolution
of a 10m-class telescope in space, such as the LUVOIR-A (15m) and LUVOIR-B (8m) mission concepts,
is needed \citep{Gal16}. In particular:

$\bullet$~ The outstanding spatial resolution of a nearly difracction-limited
$\sim$~10m telescope in space to disentangle 30~Dor-like concentrations
of massive stars out to 1~Mpc. Coupled to
integral-field spectrographs (optimized after lessons learnt from JWST), they will provide unprecedented
constraints on the IMF of both dense, and starburst regions.

$\bullet$~ High resolution ($R$=$\lambda/\Delta/\lambda \geq$10,000)
multi-object optical and near-IR spectroscopy to constrain stellar parameters.
The Extremely Large Telescope will have superior sensitivity
and reach the faintest objects,
but continuous UV-optical-NIR coverage in the
nearly background-free space environment
is still highly desirable for reliable flux calibration.

$\bullet$~ Ultraviolet multi-object spectroscopy with
medium resolution ($R$=$\lambda/\Delta/\lambda \geq$5,000)
to confirm the presence of winds, and to constrain mass loss rates and velocity fields.
In this respect the multi-object capabilities of LUMOS
coupled with on-going improvements on UV detectors,
will revolutionize the
field, by enabling the first extensive characterization of the outflows of massive stars beyond the SMC.

The selection of the most ambitious design LUMOS-A would enable studies
of individual stars in the sparse regions of I~Zw18
which currently stands the reference of extremely metal-poor, starburst dwarf galaxies.
However, both LUMOS-A and LUMOS-B will comfortably reach
out to few Mpc distances, opening great discovery opportunities of even metal-poorer massive stars
in the Sculptor, Centaurus and M81 Groups.
On the whole, both designs will provide definitive answers to the most
pressing questions in our understanding of the properties and evolution of high-mass
stars in very metal poor environments.

\pagebreak

\end{document}